%Paper: hep-ph/9302228
%From: HEHTH@SLACVM.SLAC.Stanford.EDU
%Date: Fri, 05 Feb 1993   16:44 -0800 (PST)

%macropackage=phyzzx

%\unlock
%\def\refitem#1{\r@fitem{[#1]}}
%\lock
\def\({[}
\def\){]}
\def\ifmath#1{\relax\ifmmode #1\else $#1$\fi}
\def\half{\ifmath{{\textstyle{1 \over 2}}}}
%% "low subscript" %%
\def\ls#1{\ifmath{_{\lower1.5pt\hbox{$\scriptstyle #1$}}}}
\def\rta{\rightarrow}
\def\mhl{m_{h^0}}
\def\mhh{m_{H^0}}
\def\mha{m_{A^0}}
\def\mhc{m_{H^\pm}}
\def\hl{h^0}
\def\hh{H^0}
\def\ha{A^0}

\def\mz{m_Z}
\def\M{{\cal M}}
\def\rta{\rightarrow}
\def\crr{\crcr\noalign{\vskip .15in}}
\def\bold#1{\setbox0=\hbox{$#1$}%
     \kern-.025em\copy0\kern-\wd0
     \kern.05em\copy0\kern-\wd0
     \kern-.025em\raise.0433em\box0 }

\def\Weizmann{\centerline{\it Weizmann Institute of Science}
  \centerline{\it Physics Department, Rehovot 76100, Israel}}
\def\Scipp{\centerline{\it Santa Cruz Institute for Particle Physics}
  \centerline{\it University of California, Santa Cruz, CA 95064}}
%{\baselineskip=11pt
\Pubnum={SCIPP-92/42 \cr WIS-92/61/Dec-PH \cr Revised Version}
\date={January, 1993}
\titlepage
\vbox to 1cm{}
\title{{\bf $\bold{Z\rarrow A^0A^0\nu\bar\nu}$ and
$\bold{e^+e^-\rta\ha\ha Z}$ in Two Higgs Doublet Models}}
\vskip12pt
\centerline{Howard E. Haber
\foot{Supported in part by the Department of Energy.} }
\Scipp
\vskip6pt
\centerline{Yosef Nir
\foot{Incumbent of the Ruth E. Recu Career Development Chair
supported in part by the Israel Commission for Basic Research,
by the United States---Israel Binational Science Foundation,
and by the Minerva Foundation.}}
\Weizmann

\vskip 1cm

\centerline{\bf Abstract}
In LEP searches for the neutral CP-odd scalar $A^0$ of a
multi-Higgs doublet model, experimenters have searched for $Z\rta\hl\ha$
(where $\hl$ is the lightest CP-even scalar).
No model-independent limit on the $\ha$ mass can be deduced from
present data if $\mhl>\mz$.  In this paper,
we compute the rates for $Z\rarrow A^0A^0\nu\bar\nu$
and $e^+e^-\rta \ha\ha Z$.   Assuming that no light CP-even
neutral scalars exist,the branching ratio for $Z\rta\ha\ha\nu\bar\nu$
is found to be less than $1.4\times10^{-8}$.  At higher $e^+e^-$
center-of-mass energies, $\sigma(e^+e^-\rta\ha\ha Z)$ peaks at $0.5$~fb.
A comparison with other direct searches for the $\ha$
is briefly considered.
\vfill
\centerline{Submitted to {\sl Physics Letters B}}
\endpage
%}

\titlestyle{1. Introduction}
\smallskip

\REF\hhg{For a detailed review of the two-Higgs doublet model and
a guide to the literature, see Chapter 4 of: J.F. Gunion, H.E. Haber,
G.L. Kane and S. Dawson, {\it The Higgs Hunter's Guide}
(Addison-Wesley Publishing Company, Reading, MA, 1990).}
The simplest extension of the scalar sector of the Standard Model
is a model with two complex scalar Higgs doublets (with a
CP-conserving Higgs sector)\refmark\hhg.
This model possesses
five physical scalars: two neutral CP-even scalars ($h^0$ and $H^0$
with $\mhl\leq\mhh$), one neutral CP-odd scalar ($A^0$)
and a charged pair of scalars ($H^\pm$).
In a generic (CP-conserving) two-Higgs-doublet model, the four
masses $\mhl$, $\mhh$, $\mha$ and $\mhc$
are independent parameters.  Moreover, the scalar couplings to vector
bosons and fermions depend on two additional independent parameters
of the model: the ratio of vacuum expectation values ($\tan\beta$)
and the CP-even Higgs mixing angle ($\alpha$).
This makes the experimental search for the various Higgs scalars
a complicated task.

\REF\aleph{D. Decamp \etal\ [ALEPH Collaboration], {\sl Phys. Rep.}
{\bf 216} (1992) 253.}
\REF\lthree{O. Adriani \etal\ [L3 Collaboration], {\sl Phys. Lett.}
{\bf B294} (1992) 457.}
\REF\pdg{K. Hisaka \etal\ [Particle Data Group], {\sl Phys. Rev.}
{\bf D45} (1992) S1.}
\REF\sally{S. Dawson, {\sl Nucl. Phys.} {\bf B339}
(1990) 19; Markus E. Lautenbacher, {\sl Nucl. Phys.} {\bf B347}
(1990) 120.}
\REF\crystal{D. Antreasyan \etal, {\sl Phys. Lett.} {\bf B251} (1990)
204.}
The fact that the charged scalars have not been observed in $Z$ decays
puts an unambiguous lower bound on their masses,
$\mhc\geq41.7$~GeV\refmark{\aleph,\lthree}.
The Higgs search is more complicated in the case of the neutral scalars.
The $Z$-boson does not couple to a pair of identical scalars
but only to $h^0A^0$ and $H^0A^0$.  The relative coupling strengths
are $\cos(\beta-\alpha)$ and $\sin(\beta-\alpha)$ respectively,
which in the general two-doublet model is independent of the Higgs
masses.
Thus, if $\mhl\geq\mz$ or if $|\cos(\beta-\alpha)|\ll 1$ and
$\mhh\geq\mz$, then associated production of $\ha$ in $Z$ decay
could not have been observed even if $\ha$ were massless!
Experimental limits on $\ha$ that do not depend on the
other Higgs mass parameters
of the theory are meager.  There are no official
limits in the 1992 compilation of the Particle Data Group\refmark\pdg.
Presumably, combining results from searches for $K\rta\pi\ha$,
$B\rta K\ha$, $\psi\rta\ha\gamma$ and $\Upsilon\rta\ha\gamma$ would
rule out very light CP-odd scalars\refmark{\sally,\crystal}.
However, no general $\mha$ bound above a few GeV exists today.

\REF\hhgii{See \S 4.2 of ref.~[1].}
\REF\gutu{J.F Gunion and A. Turski, {\sl Phys. Rev.} {\bf D39} (1989)
2701;
{\bf D40} (1989) 2333;
M.S. Berger, {\sl Phys. Rev.} {\bf D41} (1990) 225.}
\REF\leter{H.E. Haber and R. Hempfling, {\sl Phys. Rev. Lett.} {\bf 66}
(1991) 1815; SCIPP-91/33 (1992);
Y. Okada, M. Yamaguchi and T. Yanagida,
{\sl Prog. Theor. Phys.} {\bf 85} (1991) 1;
{\sl Phys. Lett.} {\bf B262} (1991) 54;
J. Ellis, G. Ridolfi and F. Zwirner, {\sl Phys. Lett.}
{\bf B257}, 83 (1991); {\bf B258} (1991) 395.}
\REF\other{R. Barbieri, M. Frigeni and F. Caravaglios
{\sl Phys. Lett.} {\bf B258} (1991) 167;
A. Yamada, {\sl Phys. Lett.} {\bf B263} (1991) 233;
P.H. Chankowski, S. Pokorski and J. Rosiek, {\sl Phys. Lett.}
{\bf B274} (1992) 191; {\bf B281} (1992) 100;
{\bf B286} (1992) 307; A. Brignole,
{\sl Phys. Lett.} {\bf B281} (1992) 284;
D.M. Pierce, A. Papadopoulos and S. Johnson,
{\sl Phys. Rev. Lett.} {\bf 68} (1992) 3678; K. Sasaki, M.
Carena and C.E.M Wagner, {\sl Nucl. Phys.} {\bf B381} (1992) 66.}
\REF\ellis{A. Brignole, J. Ellis, G. Rudolfi and F. Zwirner,
\sl Phys. Lett. \bf B271 \rm (1991) 123;
[E: \bf B273 \rm (1991) 550];
A. Brignole, {\sl Phys. Lett.} {\bf B277} (1992) 313;
M.A. D\'\i az and H.E. Haber, \sl Phys. Rev. \bf D45\rm
(1992) 4246.}
\REF\lep{D. Buskulic \etal\ [ALEPH Collaboration], {\sl Phys. Lett.}
{\bf B285} (1992) 309; D. Decamp \etal\ [ALEPH Collaboration],
{\sl Phys. Lett.} {\bf B265} (1991) 475; P. Abreu \etal\
[DELPHI Collaboration], {\sl Nucl. Phys.} {\bf B373} (1992) 3;
M.Z. Akrawy \etal\ [OPAL Collaboration], {\sl Z. Phys.} {\bf C49}
(1991) 1.}
Sharper bounds on $\mha$ do exist
in the minimal supersymmetric extension of the Standard Model
(MSSM).  The Higgs sector of the MSSM is a two-Higgs-doublet model
in which the Higgs parameters are highly constrained\refmark\hhgii.
As a result, one finds tree-level
relations among the various Higgs sector parameters which
reduce the six independent parameters mentioned above to two
independent parameters.
These relations are modified when
radiative corrections are incorporated\refmark{\gutu-\ellis}.
Nevertheless, the
MSSM constraints permit a much more powerful set of experimental
bounds to be obtained by the non-observation of $Z\rta\hl\nu\bar\nu$
and $Z\rta\hl\ha$.  The present LEP bound on $\ha$ in the context
of the MSSM is $\mha\geq 20$~GeV (assuming $\tan\beta\gsim
1$)\refmark{\lthree,\lep}.
\par
In this paper, we consider the possibility of excluding a
light $\ha$ in the general (CP-conserving) non-supersymmetric
two-Higgs-doublet model.  We examine the extreme case in which
the CP-odd neutral
scalar $A^0$ is light, but all other scalars are heavy (say, with masses
above the $e^+e^-$ collider center-of-mass energy).%
\foot{The case where $\mha\ll\mz$ and
all other Higgs states are heavier than the $Z$ is not possible in the
MSSM.}  We then pose the following question: is it possible to find
(or exclude) the $\ha$ in a completely model-independent way?
Since the other Higgs scalars are assumed to be heavy, the $\ha$ must be
produced either through its couplings to fermions or vector bosons.
The Higgs-fermion interaction is model-dependent; the $\ha f\bar f$
coupling strength depends on $\tan\beta$.  On the other hand, the
four-point couplings of $\ha\ha$ to $W^+W^-$ and $ZZ$ and their
respective coupling strengths are fixed by the gauge invariance of the
theory.  Thus, we shall restrict our attention in this
paper to $\ha$ production processes that only involve the
$\ha$ couplings to vector bosons.

\FIG\figA{Feynman diagram for $Z\rarrow A^0A^0\nu\bar\nu$.
In the limit where all other Higgs scalars are much heavier than
$\mz$, other contributing tree-level diagrams (mediated by
virtual heavy Higgs exchange) can be neglected.}

%5\topinsert
%   \tenpoint \baselineskip=12pt   \narrower
%\plotpicture{\hsize}{5cm}{aafig.topdraw}
%\vskip12pt\noindent
%{\bf Fig.~\figA.}\enskip
%Feynman diagram for $Z\rarrow A^0A^0\nu\bar\nu$.
%In the limit where all other Higgs scalars are much heavier than
%$\mz$, other contributing tree-level diagrams (mediated by
%virtual heavy Higgs exchange) can be neglected.
%
%\endinsert

In section 2, we study the decay mode $Z\rarrow A^0A^0\nu\bar\nu$
shown in Fig.~\figA.
The $\nu\bar\nu$ pair arises from decay products of a virtual
$Z$-boson. Our results can be easily extended to other
pairs of light fermions arising from the virtual $Z$.
If the heavier Higgs bosons of
the model are sufficiently heavy, then the Feynman diagram of
Fig.~\figA\ is the only relevant tree-level diagram.
If $\mha=0$, we can compute the decay rate explicitly by
analytically integrating over the four-body phase space.  This
provides an upper limit for the decay rate.  We have also carried out
the more general computation for $\mha\neq 0$ where the phase space
integration must be carried out numerically.
In section 3, we compute the cross section for $e^+e^-\rta\ha\ha Z$.
This process proceeds via a diagram similar to the one shown in
fig.~\figA, except that the roles of the real $Z$ and virtual $Z^\ast$
are reversed.  The end results of both calculations are disappointing;
the respective branching ratio and cross-sections are too small to
be detected at LEP (and LEP-200).
Finally, in section 4, we compare our result with other
direct searches for the $\ha$.
\bigskip
\titlestyle{2. Computation of $\Gamma(Z\rta\ha\ha\nu\bar\nu)$}
\smallskip

In this section, we compute the contribution of Fig.~\figA\ to
$Z\rta\ha\ha\nu\bar\nu$.  The rate depends on the
$ZZAA$ coupling which in the two-Higgs doublet model is given by
$$g\ls{Z_\mu Z_\nu A^0A^0}={ig^2\over2\cos^2\theta_W}g_{\mu\nu}\,.
\eqn\aab$$
Note in particular that this coupling does not depend on the Higgs
sector parameters $\alpha$ and $\beta$.  Thus, the decay rate
we compute will depend only on one unknown parameter---the mass of
the $\ha$.

The amplitude for $Z(p)\rta\ha(k_1)\ha(k_2)\nu(k_3)\bar\nu(k_4)$
(with the four-momenta indicated in parentheses) in unitary gauge is
$$i\M={ig^2\over2\cos^2\theta_W}\epsilon^\mu(p)
{-i\left(g_{\mu\nu}-\kappa_\mu \kappa_\nu/ m_Z^2\right)
\over \kappa^2-m_Z^2}
\bar u(k_3)\left(
{-ig\over4\cos\theta_W}\right)\gamma^\nu(1-\gamma_5)v(k_4)\,,
\eqn\aac$$
where $\kappa\equiv k_3+k_4$ is the four-momentum of the virtual $Z$.
The decay width is then
$$\eqalign{\Gamma&(Z\rarrow \ha\ha\nu\bar\nu)=
{g^6\over 96 (2\pi)^8 \mz^3\cos^6\theta_W}
\int{d^3k_3\over2E_3}{d^3k_4\over2E_4}{1\over(2k_3\cdot k_4-m_Z^2)^2}\crr
\times&\int{d^3k_1\over2E_1}{d^3k_2\over2E_2}
\left\(m_Z^2(k_3\cdot k_4)+2(k_3\cdot p)(k_4\cdot p)\right\)
\delta^{(4)}(p-k_1-k_2-k_3-k_4)\,,\cr}\eqn\aad$$
where we have included a factor of ${1\over 2}$ to account for identical
$\ha$ bosons in the final state and a factor of ${1\over 3}$ for the
$Z$ spin-average.
\REF\lackner{See Appendix A of Y. Singh, {\sl Phys. Rev.} {\bf 161}
(1967) 1497, and Appendix D of R.M. Barnett, H.E. Haber and
K.S. Lackner, {\sl Phys.~Rev.} {\bf D29} (1984) 1990.}
To integrate over $d^3k_1d^3k_2$ we use\foot{Here, we follow the
techniques of ref.~[\lackner].}
$$\int{d^3k_1\over2E_1}{d^3k_2\over2E_2}
\delta^{(4)}(N-k_1-k_2)\times\cases{
1&$={\pi\over2}R_4^{1/2}$,\cr
k_{1\mu}&$={\pi\over4}R_4^{1/2}N_\mu$,\cr
k_{1\mu}k_{1\nu}&$=-{\pi\over24}R_4^{1/2}
(R_4N^2g_{\mu\nu}-4R_1N_\mu N_\nu)$,\cr
k_{1\mu}k_{2\nu}&$={\pi\over24}R_4^{1/2}
(R_4N^2g_{\mu\nu}+2R_{-2}N_\mu N_\nu)$,\cr
}\eqn\aae$$
where $N$ is an arbitrary four-vector and
$$R_n\equiv1-{n\mha^2\over N^2}\,.\eqn\aaf$$
Applying this result to eq.~\aad, we find
$$\eqalign{\Gamma(Z\rarrow AA\nu\bar\nu)=
{g^6\over3(4\pi)^7\mz^3\cos^6\theta_W}
\int&{d^3k_3\over2E_3}{d^3k_4\over2E_4}
\left(1-{4\mha^2\over(p-k_3-k_4)^2}\right)^{1/2}\crr \times&
{m_Z^2(k_3\cdot k_4)+2(k_3\cdot p)
(k_4\cdot p)\over\(2(k_3\cdot k_4)-m_Z^2\)^2}.\cr}\eqn\aag$$

It is convenient to
work in the $Z$ rest frame:
$$p=(m_Z;0,0,0);\ \  k_3=E_3(1;0,0,1);\ \
k_4=E_4(1;\sin\theta,0,\cos\theta),\eqn\aah$$
where $E_3$ and $E_4$ are the neutrino energies.  We introduce
the following scaled kinematic variables
$$v\equiv{4\mha^2\over m_Z^2},\ \ \ w\equiv{1-\cos\theta\over2},\ \ \
y\equiv{2E_3\over m_Z},\ \ \ z\equiv{2E_4\over m_Z}\,.\eqn\aai$$
Then
$$\eqalign{\Gamma(Z\rarrow AA\nu\bar\nu) &=
{g^6 m_Z\over3\cdot2^{17}\pi^5\cos^6\theta_W}
\int_0^{1-v}\!dz\! \left\(\int_0^{1-v-z}dy\int_0^1 dw+\!
\int_{1-v-z}^{1-v}dy\! \int_{(y+z-1+v)/yz}^1 \!dw\right\)\crr%
&\qquad\qquad\times
{y^2z^2(1+w)\over(yzw-1)^2}\left(1-{v\over1-y-z+yzw}\right)^{1/2}
\,.\cr}\eqn\aaj$$

For a massless $A^0$ ($v=0$), the square-root factor in eq.~\aaj\ is
unity, and the remaining integrals can be carried out explicitly.
The result is
$$\Gamma(Z\rarrow AA\nu\bar\nu)=
{g^6 m_Z\over3\cdot2^{17}\pi^5\cos^6\theta_W}
\left\({\pi^2\over2}-{349\over72}\right\).\eqn\aak$$
It is unfortunate that the factor in brackets is rather small
(roughly 0.09), thereby reducing
the partial width by more than an order of magnitude over
a naive initial estimate.  The branching ratio is most easily
displayed by normalizing our result to $\Gamma(Z\rta\nu\bar\nu)=
g^2\mz/(96\pi\cos^2\theta_W)$.  We find
$${\Gamma(Z\rta\ha\ha\nu\bar\nu)\over\Gamma(Z\rta\nu\bar\nu)}=
{\alpha^2({\pi^2\over 2}-{349\over 72})\over 256\pi^2\sin^4\theta_W
\cos^4\theta_W}\,,\eqn\aaextra$$
where $\alpha\simeq 1/128$ is the fine-structure constant evaluated
at the $Z$ mass.  Noting that the branching ratio for $Z$ decay to
three generations of neutrinos is $20.2\%$, we end up with
$$BR(Z\rarrow A^0A^0\nu\bar\nu)=1.4\times10^{-8}.\eqn\aal$$

\FIG\figB{Phase space suppression of $Z\rta\ha\ha\nu\bar\nu$ as a
function of $\mha$.  $\Gamma_0$ is the corresponding
decay rate for $\mha=0$; at this point,
$BR(Z\rarrow A^0A^0\nu\bar\nu)=1.4\times10^{-8}.$}
%\topinsert
%   \tenpoint \baselineskip=12pt   \narrower
%\plotpicture{\hsize}{9cm}{aa.topdraw}
%\vskip12pt\noindent
%{\bf Fig.~\figB.}\enskip
%Phase space suppression of $Z\rta\ha\ha\nu\bar\nu$ as a
%function of $\mha$.  $\Gamma_0$ is the corresponding
%decay rate for $\mha=0$; at this point,
%$BR(Z\rarrow A^0A^0\nu\bar\nu)=1.4\times10^{-8}$.
%
%\endinsert
For $\mha\neq0$
we have carried out the integrations in eq.~\aaj\ numerically.
The results are presented in Fig. \figB.
The branching ratio decreases from the result of eq.~\aal\
by about an order of magnitude
for $\mha\sim15$~GeV and by two orders of magnitude for
$\mha\sim25$~GeV.
\bigskip
\titlestyle{3. $e^+e^-\rta \ha\ha Z$ at a
Higher Energy $e^+e^-$ Collider}
\smallskip

In light of the extremely small branching ratios for $Z\rta\ha\ha\nu
\bar\nu$ obtained above, we extend our considerations to
future Higgs searches at higher energy $e^+e^-$ colliders
(LEP-200 and the next $e^+e^-$ linear collider (NLC) with
$\sqrt{s}\geq 300$~GeV).  Consider first the search for the lightest
CP-even Higgs ($\hl$) at LEP. If $\hl$ is not discovered
at LEP-I (say, via $Z\rta\hl\nu\bar\nu$), then the search will be
continued at LEP-II via $e^+e^-\rta\hl Z$.  Once the light
CP-even Higgs state is found, one can, in principle, place
unambiguous bounds on the existence of $\ha$ if $e^+e^-\rta\hl\ha$
is not observed.  On the other hand, suppose that $\hl$ is still
not discovered.  Is it possible that a rather light $\ha$ could
still elude experimental detection?

\REF\sher{H.E. Haber and M. Sher, {\sl Phys. Rev.} {\bf D35} (1987)
2206; M. Drees, {\sl Phys. Rev.} {\bf D35} (1987) 2910;
J.R. Espinosa and M. Quiros, {\sl Phys. Lett.} {\bf B279} (1992) 92; G.L.
 Kane, C. Kolda and J.D. Wells, Michigan preprint UM-TH-92-24 (1992).}
\REF\lane{See, \eg, K. Lane, in {\it Proceedings of the 1982 DPF
Snowmass
Summer Study on Elementary Particle Physics and Future Facilities},
edited by R. Donaldson, R. Gustafson, and F. Paige (Fermilab, Batavia,
Illinois, 1982), p.~222.}
It might be argued that it is theoretically unlikely that
$\ha$ is light while the lightest CP-even scalar
is quite heavy.  This prejudice is common among
advocates of low-energy supersymmetry models\refmark\sher.
In contrast, light CP-odd and heavy CP-even scalars are natural
in technicolor models\refmark\lane.  For example, many technicolor models
possess a spectrum of light CP-odd pseudo-Goldstone bosons (``pseudos'');
those states that are electrically and color-neutral
would behave like the
$\ha$ of the two-doublet Higgs model.  In contrast,
the CP-even scalars would
be spin 0 bound states of techni-fermions and would almost certainly
be considerably heavier than the light pseudos.  Thus, until a
CP-even Higgs scalar is discovered, it will be important
to study methods for discovering the $\ha$ in processes that do not
involve the CP-even scalars.

In analogy with the search for the CP-even scalar at LEP-200, we
consider the mechanism for $\ha\ha$ production shown
in fig.~\figA, where the role of the $Z$ and $Z^\ast$ are
interchanged.  That is, we consider $e^+e^-\rta \ha\ha Z$ via
$s$-channel $Z^\ast$-exchange.  Here, we assume that all
other Higgs scalars are much heavier than the $e^+e^-$
center-of-mass energy $\sqrt{s}$ so that other
contributing tree-level diagrams mediated by the heavier CP-even scalars
can be omitted.  In this case, only the one diagram contributes.
The squared matrix element for $e^+e^-\rta \ha\ha Z$,
averaged (summed) over initial (final) spins is given by:
$$|{\cal M}|^2_{\rm ave}={g^6[1-4\sin^2\theta_W+8\sin^4\theta_W]\over
64\mz^2\cos^6\theta_W(s-\mz^2)^2}\,\left[\mz^2 s+4p_1\cdot k\ls{Z}\,
p_2\cdot k\ls{Z}\right]\,,\eqn\sqme$$
where $p_1$, $p_2$ and $k\ls{Z}$ are the $e^-$, $e^+$, and $Z$
four-momenta respectively.
In evaluating the cross-section for $e^+e^-\rta \ha\ha Z$,
we make use of eq.~\aae\ in the integration over the three-body phase
space.  This leaves one last integral over the three-momentum of the
$Z$.  At this point it is convenient to work in the $e^+e^-$
center-of-mass frame.
\REF\kinematics{E. Byckling and K. Kajante, {\it Particle Kinematics},
(John Wiley \& Sons, New York, 1973).}
In this frame, the energy of the $Z$
is $E\ls{Z}=(s+\mz^2-s_2)/2\sqrt{s}$ where $s_2$ is the invariant mass
of the $\ha\ha$ pair recoiling against the
$Z$\refmark\kinematics.  From $4\mha^2\leq s_2\leq (\sqrt{s}-\mz)^2$,
we easily obtain the limits of integration for the variable $E\ls{Z}$.
Introducing the scaled variables:
$$v\equiv{4\mha^2\over s},\ \ \
x\equiv{\mz\over\sqrt{s}},\ \ \ y\equiv{E\ls{Z}\over\sqrt{s}}
\,,\eqn\scaled$$
we end up with:
$$\eqalign{\sigma(e^+e^-\rta \ha\ha Z)&={g^6 s^2 [1-4\sin^2\theta_W
+8\sin^4\theta_W]\over 3\cdot2^{12}\pi^3\cos^6\theta_W \mz^2
(s-\mz^2)^2}\cr \times&\int_x^{\half(1+x^2-v)}\,dy\,(y^2-x^2)^{1/2}
(y^2+2x^2)\left(1-{v\over 1+x^2-2y}\right)^{1/2}\,.\cr}\eqn\sigzaa$$
In the above formula, we have already integrated over the angular
direction of the $Z$, and we have included the factor of
$\half$ due to identical $\ha$ bosons in the final state.
Once again, we can perform the final integral analytically when
$\mha=0$.  The result is
$$\eqalign{\sigma(e^+e^-\rta \ha\ha Z)&={g^6 s^2 [1-4\sin^2\theta_W
+8\sin^4\theta_W]\over 3\cdot2^{18}\pi^3\cos^6\theta_W \mz^2
(s-\mz^2)^2}\crr&\quad\times \left[(1-x^4)(1+16x^2+x^4)+72x^4\ln x\right]
\,.\cr}\eqn\sigzaazero$$

It is perhaps surprising that in the limit of large $s$, these
cross-sections approach a constant value rather than behaving like
$1/s$ as expected for annihilation cross-sections.  This behavior
arises from the production of a longitudinal $Z$ in the final state.
Moreover,
had we included the other tree-level contributions involving
the exchange of the CP-even scalars (both $\hl$ and $\hh$), we would
find that the leading ${\cal O}(1)$ behavior of the cross-section
cancels exactly, leaving the expected ${\cal O}(1/s)$ behavior.
We have checked explicitly that this cancellation does occur.
This means that eq.~\sigzaa\ and \sigzaazero\ can only be accurate
for $\sqrt{s}$ smaller than the CP-even scalar masses; for larger
values of $\sqrt{s}$, these formulae must overestimate the true results.
\FIG\figC{Cross-section for $e^+e^-\rta \ha\ha Z$.  We exhibit curves
in two cases: (a) as a function of $\sqrt{s}$ for
$\mha=0$, 20 and 40~GeV and (b) as
a function of $\mha$ for $\sqrt{s}=150$, 175, 200 and 500~GeV.}
%\topinsert
%   \tenpoint \baselineskip=12pt   \narrower
%\plotpicture{\hsize}{9cm}{zaafig3.topdraw}
%\vskip12pt\noindent
%{\bf Fig.~\figC.}\enskip
%Cross-section for $e^+e^-\rta \ha\ha Z$.  We exhibit curves
%in two cases: (a) as a function of $\sqrt{s}$ for
%$\mha=0$, 20 and 40~GeV and (b) as
%a function of $\mha$ for $\sqrt{s}=150$, 175, 200 and 500~GeV.
%
%\endinsert

We now turn to the numerical implications of the above results.
In fig.~\figC, we plot the cross-section for $e^+e^-\rta \ha\ha Z$
as a function of the center-of-mass energy $\sqrt{s}$ for three choices
of $\mha$.  We also exhibit the dependence on $\mha$ for possible
energies
of LEP-200 and the NLC (with $\sqrt{s}=500$~GeV).  The cross-sections
are disappointing.  At LEP-200, optimistic projections suggest
yearly data accumulations of about $500~{\rm pb}^{-1}$.  This would
imply less than one $\ha\ha Z$ event per year, even before detection
efficiency and $Z$ branching ratio factors are taken into account.
\REF\saariselka{B. Wiik, in {\it Physics and Experiments with
Linear Colliders} (Volume I), Workshop Proceedings,
Saariselk\"a, Finland, 9-14 September 1991, edited by R. Orava,
P. Eerola and M. Nordberg (World Scientific, Singapore, 1992)
p.~83. See also the contributions of B. Richter (p.~611), R. Ruth
(p.~629), T. Weiland (p.~663), M. Tigner (p.~705), and G.-A. Voss
(p.~777) in {\it op.~cit.} (Volume II).}
At the NLC, typical design luminosities yield between 20 and
$50~{\rm fb}^{-1}$ of data per year\refmark\saariselka.
In this case, at most a few
$\ha\ha Z$ events per year could be detected (after
experimental triggers and cuts).
Of course, such a search would only be critical if no
$\hl$ were discovered at the NLC, as discussed at the beginning of
this section.

\REF\gunionetal{J.F. Gunion \etal, {\sl Phys. Rev.} {\bf D38} (1988)
3444.}
Finally, for completeness, we note that other production processes
for $\ha\ha$ should be considered at the NLC once the center-of-mass
is sufficiently large.  For example, one
can consider $e^+e^-\rta \ha\ha W^+W^-$ and $e^+e^-\rta \ha\ha ZZ$
where the $\ha\ha$ is emitted from one of the
outgoing vector boson lines.
In addition, for $\sqrt{s}\geq 500$~GeV, vector boson fusion
processes begin to become important.  Thus, one should examine
$e^+e^-\rta\ha\ha \nu\bar\nu$ (via $W^+W^-$ fusion) and
$e^+e^-\rta\ha\ha e^+e^-$ (via $ZZ$ fusion).  The cross-sections
for all these processes
are formally of higher order [${\cal O}(g^8)$] as compared to
$e^+e^-\rta \ha\ha Z$ considered above.  However, the cross-section
for the fusion processes increase logarithmically with $\sqrt{s}$ and
eventually dominate the annihilation processes.  The $W^+W^-$ fusion
cross-section is the largest and was computed in ref.~\gunionetal.
Unfortunately, even at $\sqrt{s}=2$~TeV, its cross-section appears to
be too small for detection at a future $e^+e^-$ linear collider.
\bigskip
\titlestyle{4. Discussion and Conclusions}
\smallskip
\REF\wlodek{E. Duchovni and T. Wlodek, preprint in preparation.}
\par
The branching ratio for $Z\rta\ha\ha\nu\bar\nu$ is probably too
small to be observed even at a high luminosity $Z$ factory.
Including other possible decays of the virtual $Z$ to fermion
pairs can only enhance the branching ratio by a factor of 5.
Even with $100\%$  detection efficiency, it seems unlikely that
the $\ha$ can be detected.  Our
calculation has assumed the most general (CP-conserving)
two-Higgs-doublet model with a light $A^0$ and all other Higgs scalars
very heavy. If the other scalars are not all heavy, there are
additional diagrams that contribute to $Z\rta\ha\ha\nu\bar\nu$.
%In general, the parameter space is large.
We expect an enhancement of the
$Z\rta\ha\ha\nu\bar\nu$ decay rate to be at least an order of
magnitude beyond the result of eq.~\aal, since there is no
reason for a similar suppression by the factor $\pi^2/2-349/72$
as found in eq.~\aak.  After adding up all diagrams involving
the other Higgs states of the model, we would expect a branching ratio
perhaps as large as $10^{-6}$.  A more complete calculation (in the
context of the MSSM) is currently in progress\refmark\wlodek.
Such a decay rate may be observable
at a future high intensity $Z$ factory.  However, a favorable
branching ratio would exist only in limited regions of the parameter
space.  Going to a higher energy $e^+e^-$ collider does not
improve the picture substantially.  In this case, one would search
for $e^+e^-\rta\ha\ha Z$.  Based on our computations, we would
conclude that the production cross sections are not large enough
to permit experimental detection at LEP-200.  At an even higher
energy NLC with a data collection greater than $10~{\rm fb}^{-1}$
per year, at most a few signal events per year could be detected.
Thus, an unambiguous model-independent limit on the $\ha$
mass seems to be unattainable.

\REF\lingfong{L.-F. Li, CMU-HEP-91-10 (1991), presented at the
14th International Warsaw meeting on Elementary Particle Physics,
Warsaw, Poland, 27-31 May 1991.}
\REF\zerwas{A. Djouadi, P.M. Zerwas, and J. Zunft,
{\sl Phys. Lett.} {\bf B259} (1991) 175.}
It is instructive to compare the results of this paper with other
proposals to search for $\ha$ in rare $Z$ decays.  Two comparable
decay processes have been studied in the literature:
$Z\rta\ha\ha\ha$ and $Z\rta b\bar b\ha$.
L.-F. Li has estimated\refmark\lingfong\
the rate for $Z\rta\ha\ha\ha$ via a one-loop box diagram with a
t-quark running around the loop.  He estimates
$BR(Z\rta\ha\ha\ha)\sim 10^{-5}$, although his estimate is rather
crude.  If this estimate is not overly optimistic, then it clearly
presents an excellent chance to obtain an experimental bound
on $\mha$ which is independent of the other Higgs masses of the
model.  However, there is still some model-dependence in this case,
since the amplitude for $Z\rta\ha\ha\ha$ (via the
$t$-quark loop) depends on the $\ha t\bar t$ coupling.  This
coupling depends on $\tan\beta$ and could be suppressed.
Of course, if $\hl$ is not too heavy, then the dominant
contribution to $Z\rta\ha\ha\ha$ would arise from $Z\rta\ha h^{0*}$
in which the virtual $h^{0*}$ decays to $\ha\ha$.

The decay rate for $Z\rta b\bar b\ha$ also depends on the
model-dependent Higgs-fermion coupling.  Djouadi, Zerwas and Zunft
find $BR(Z\rta b\bar b\ha)\simeq 3\times 10^{-4}$ for $\mha=10$~GeV
and $\tan\beta=20$\refmark\zerwas.
(This branching ratio scales as $\tan^2\beta$.)
Once again, we have the potential for obtaining limits on $\ha$
independent of the masses of the other Higgs scalars of the theory.
As before, the model-dependence of the Higgs-fermion coupling still
remains.  However, if we subscribe to the Model-II Higgs-fermion
couplings of ref.~[\hhg] (this coupling pattern is the one that
appears in the MSSM),
then the $\ha$ coupling to $b\bar b$
($t\bar t$) is proportional to $\tan\beta$ ($\cot\beta$).  Thus,
the latter two processes are complimentary.  A combined search
for $Z\rta\ha\ha\ha$ and $Z\rta b\bar b\ha$ at a future
high intensity $Z$ factory could then rule out
the possible existence of a light $\ha$.
At a higher energy $e^+e^-$ collider (such as the NLC),
one can search for both $e^+e^-\rta b\bar b\ha$ and
$e^+e^-\rta t\bar t\ha$ with similar effect.
On the other hand, in the Model-I Higgs-fermion coupling of
ref.~[\hhg], it is possible that both $\ha$ couplings to $b\bar b$
and $t\bar t$ are suppressed.  This leaves open the possibility
that a light $\ha$ could escape experimental detection for the
foreseeable future.
\bigskip
\line{\fourteenrm\hfil ACKNOWLEDGEMENTS\hfil}
\bigskip
 We gratefully acknowledge the probing questions of Ehud Duchovni
 that encouraged us to examine the issues discussed in this work.
We also thank Carlos Figueroa for checking some of the
calculations presented in this paper.
%This work was supported
%in part by the Israel Commission for Basic Research,
%by the United States---Israel Binational Science Foundation,
%by the Minerva Foundation, and by the U.S. Department of Energy.
\bigskip
\refout
\endpage
\figout
\end